\documentclass[ 12pt, nonfootinbib, notitlepage]{revtex4-1}
\usepackage{amsmath}
\usepackage{amsfonts}
\usepackage{amsthm}
\usepackage{amssymb}
\usepackage{graphicx}
\usepackage{epsfig}
\usepackage{hyperref}
\usepackage{amscd}
\usepackage{xcolor}
\usepackage{enumerate}

\begin{document}
\title{Comment on ``The space-time line element for static ellipsoidal objects''. }

\author{Antonio  C. Guti\'errez-Pi\~{n}eres}
\email{acgutier@uis.edu.co}
\affiliation{Escuela de F\'\i sica, Universidad Industrial de Santander, CP 680002,  Bucaramanga, Colombia.}



\maketitle

Recently, an interesting paper \cite{brahma2023space} was published in which the authors employed a procedure similar to that presented in \cite{Chandrasekhar2002mathematical} to obtain a solution of the Einstein field equations for a class of static space‑times describing the exterior of a source with ellipsoidal symmetry. The constructed line element depends on a parameter $\eta$, interpreted as the linear eccentricity. In the special case $\eta=0$, the geometry reduces to spherical symmetry, and the metric collapses to the Schwarzschild solution.
In an erratum to their original work \cite{brahma2024correction}, the authors write the metric in the form
\begin{align}
ds^2 & =  [1 - A(r, \theta)]c^2dt^2 - B(r, \theta) \frac{r^2 dr^2}{r^2 - \eta^2\sin^2\theta} 
            -  [1 + C(r,\theta)][r^2 + \eta^2 (\cos^2\theta - \sin^2\theta)]d\theta^2 \nonumber\\
        &  - [(r^2 + \eta^2\cos^2\theta) \sin^2\theta d\phi^2
            +2 B(r, \theta) \frac{r\theta^2\sin\theta\cos\theta}{r^2 - \eta^2\sin^2\theta} dr d\theta \ , 
 \end{align}
where
\begin{align*}
&A(r, \theta)= \frac{2m(r^2 - \eta^2\sin^2\theta)^{1/2}}{r^2 + \eta^2\cos^2\theta} \ ,\
B(r, \theta)= \frac{r^2 + \eta^2(\cos^2\theta - \sin^2\theta)^{1/2}}{r^2 + \eta^2\cos^2\theta - 2m(r^2 - \eta^2\sin^2\theta)} 
\end{align*}
and
\begin{align*}
C(r, \theta)= \frac{\frac{\eta^4\sin^2\theta\cos^2\theta}{r^2 - \eta^2\sin^2\theta}}{r^2 + \eta^2\cos^2\theta - 2m(r^2 - \eta^2\sin^2\theta)} \ .
\end{align*}
To assess the validity of the authors’ claims, we implemented the proposed space‑time metric using the {\it Differential Geometry} package of {\bf Maple 2021}. Our computations reveal that both the Einstein tensor and the Ricci scalar are non‑vanishing. We also treated the line element as that of an anisotropic fluid and obtained non‑zero pressure components. Therefore, if our calculations are correct, we conclude that the metric presented by the authors does not represent the vacuum exterior of a static ellipsoidal source.

Additionally, in the Introduction of their paper, the authors cite the work \cite{nikouravan2011schwarzschild}, which, according to their discussion, presents a line element intended to describe static and ellipsoidal configurations. We have examined this metric as well and find that it likewise fails to satisfy the vacuum Einstein field equations.

%

\end{document}